\documentclass{article}
\usepackage{standardArxiv}
\usepackage[utf8]{inputenc} 
\usepackage[T1]{fontenc}    
\usepackage{hyperref}       
\usepackage{url}            
\usepackage{booktabs}       
\usepackage{amsfonts}       
\usepackage{amsmath}
\DeclareMathOperator{\Tr}{Tr}
\usepackage{nicefrac}       
\usepackage{microtype}      
\usepackage{lipsum}
\usepackage{fancyhdr}
\usepackage{tcolorbox}
\usepackage{graphicx}       
\usepackage{float}
\usepackage[percent]{overpic}
\graphicspath{{media/}}     
\usepackage{subcaption}

\title{\Large QSTToolkit: A Python Library for Deep Learning Powered Quantum State Tomography}

\author{
  \begin{tabular}[t]{c c}
    George FitzGerald & Will Yeadon \\[0.5ex]
    \texttt{gwfitzg@hotmail.com} & \texttt{will.yeadon@durham.ac.uk} \\[1.5ex]
    \multicolumn{2}{c}{\normalsize Department of Physics, Durham University, UK}
  \end{tabular}
}

\date{\today}

\begin{document}
\maketitle


\begin{abstract}
We introduce QSTToolkit, a Python library for performing quantum state tomography (QST) on optical quantum state measurement data. The toolkit integrates traditional Maximum Likelihood Estimation (MLE) with deep learning-based techniques to reconstruct quantum states. It includes comprehensive noise models to simulate both intrinsic state noise and measurement imperfections, enabling the realistic recreation of experimental data. QSTToolkit bridges TensorFlow, a leading deep learning framework, with QuTiP, a widely used quantum physics toolbox for Python. This paper describes the library's features, including its data generation capabilities and the various QST methods implemented. QSTToolkit is available at \url{https://pypi.org/project/qsttoolkit/}, with full documentation at \url{https://qsttoolkit.readthedocs.io/en/latest/}.
\end{abstract}

\keywords{Quantum State Tomography \and QSTToolkit \and Deep Learning \and Optical Quantum States}

\section{Introduction}
\begin{figure}[ht!]
\centering
\includegraphics[width=12.5cm]{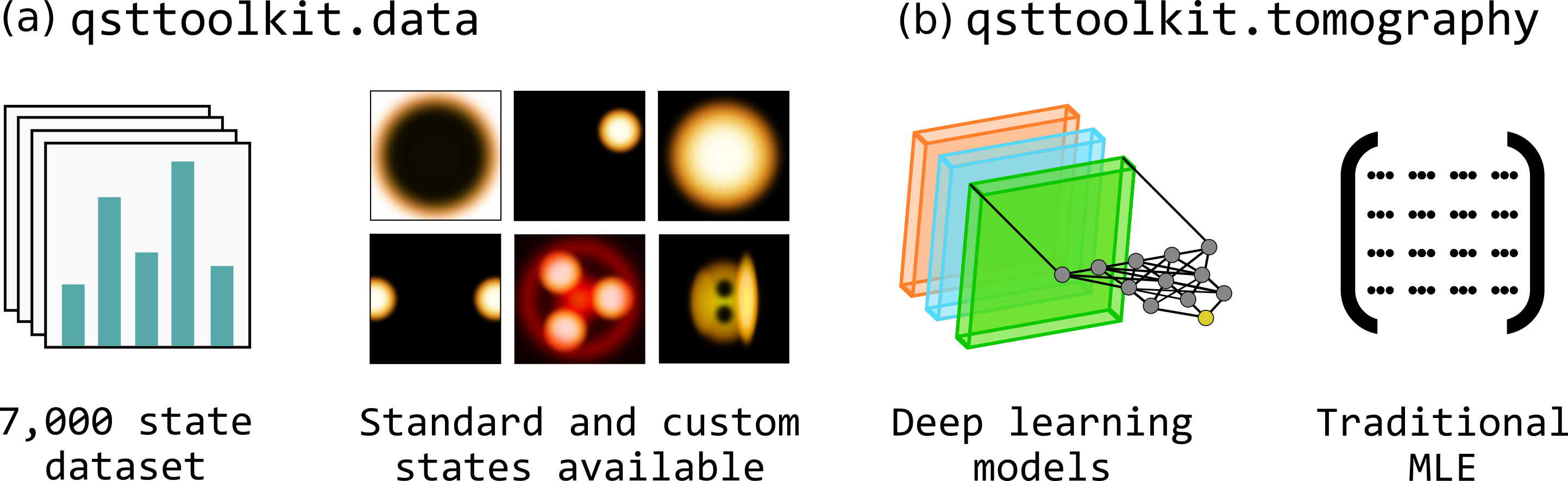}
\caption{Overview of QSTToolkit. (a) The data module provides direct access to QuTiP-generated Fock, coherent, and thermal states, as well as custom states such as Binomial and GKP. (b) The tomography module supports both traditional Maximum Likelihood Estimation (MLE) and deep-learning-based methods. To facilitate comparison between different reconstruction methods, a standard dataset containing 7000 quantum states is included.}
\label{fig-overview-alt}
\end{figure}

Quantum state tomography (QST) is the process of reconstructing a quantum state's complete mathematical description from repeated measurements of identically prepared systems~\cite{lvovsky2009}. As quantum devices scale in complexity, accurate and efficient QST becomes essential for verifying device performance, performing error correction, and guiding quantum algorithm development~\cite{zihao2024}. Traditional QST methods - including Maximum Likelihood Estimation (MLE) \cite{MLE_1}, linear inversion \cite{LI_1}, and Bayesian inference \cite{bayes_1} - can struggle with the challenges posed by high-dimensional and noisy systems. In contrast, recent advances in deep learning have provided new avenues for robust state reconstruction and classification~\cite{ahmed2021classification}. Although popular Python libraries such as QuTiP~\cite{qutip} and Qiskit~\cite{Qiskit} may offer frameworks for representing quantum states computationally, and even performing MLE or linear inversion tomography, they generally do not include realistic noise simulations or native neural network tomography capabilities. To address these challenges, we present QSTToolkit (see Figure \ref{fig-overview-alt}), a unified Python framework that integrates QuTiP's simulation capabilities with TensorFlow's deep learning functionality~\cite{tensorflow}.

The toolkit generates synthetic optical quantum state data with realistic optional noise models and implements multiple reconstruction techniques. Its design is modular, with two principal subpackages: \texttt{qsttoolkit.data} for state and measurement data generation, and \texttt{qsttoolkit.tomography} for reconstruction algorithms. The package includes both conventional statistical tomography methods and novel deep learning approaches in a common framework allowing their direct comparison. This approach enables workflows such as performing MLE on a dataset and, with the same toolkit, applying a neural network model to the same data.


\section{Key Features}
\subsection{Data Module}
\subsubsection{Optical Quantum States}
Optical quantum states in QSTToolkit are constructed using QuTiP’s \texttt{Qobj} class, with extended functionality for simulating realistic experimental conditions. The toolkit supports the generation of several canonical state types. Fock states, denoted as \(\lvert n \rangle\), represent fixed photon number states and are generated via QuTiP’s \texttt{fock(N, n)} function, where \(N\) is the Hilbert space dimension and \(n\) is the specific photon number. Coherent states, which describe displaced states in phase space, are characterized by the complex amplitude \(\alpha\) and exhibit Poissonian photon number statistics; these are created using \texttt{coherent(N, alpha)}. Thermal states, which are mixed states with super-Poissonian photon statistics, can be generated using QuTiP’s \texttt{thermal\_dm(N, nth)}, where \texttt{nth} specifies the mean thermal occupancy. 

In addition to these foundational states, QSTToolkit implements custom superpositions of Fock states, such as binomial states—formed using binomial distributions in the Fock basis and numerically optimized (“num”) states, which are tailored for specific quantum error correction protocols \cite{numbinomial}. The toolkit also provides support for cat states, which are coherent state superpositions of the form

\begin{equation}
\lvert \text{cat}_{\alpha,\pm}\rangle = \frac{1}{\sqrt{\mathcal{N}_\pm}}\bigl(\lvert \alpha \rangle \pm \lvert -\alpha \rangle \bigr),
\end{equation}

with normalization factor \(\mathcal{N}_\pm\). Even and odd cat states are particularly relevant in error correction due to their resilience against certain types of noise. Additionally, QSTToolkit provides Gottesman-Kitaev-Preskill (GKP) states, which approximate grid-like structures in phase space and serve as a foundation for fault-tolerant continuous-variable quantum computing \cite{Campagne_Ibarcq_2020}. In practice, these states are often constructed by superimposing squeezed states to form a periodic lattice in the quadrature domain. For ease of experimentation, the toolkit provides batch generators (e.g., \texttt{qsttoolkit.data.CatStates}) which allow the creation of large datasets with randomized parameters. A typical usage example is:

\begin{tcolorbox}[colback=gray!5!white,colframe=gray!75!black,title=Batch Data Generation Example]
\begin{verbatim}
import qsttoolkit as qst

# Generate 1000 cat states with randomized displacement magnitudes
cat_batch = qst.CatStates(n_states=1000, dim=dim, alpha_magnitude_range=[0,10])
\end{verbatim}
\end{tcolorbox}

\begin{figure}
    \centering
    \includegraphics[width=.75\linewidth]{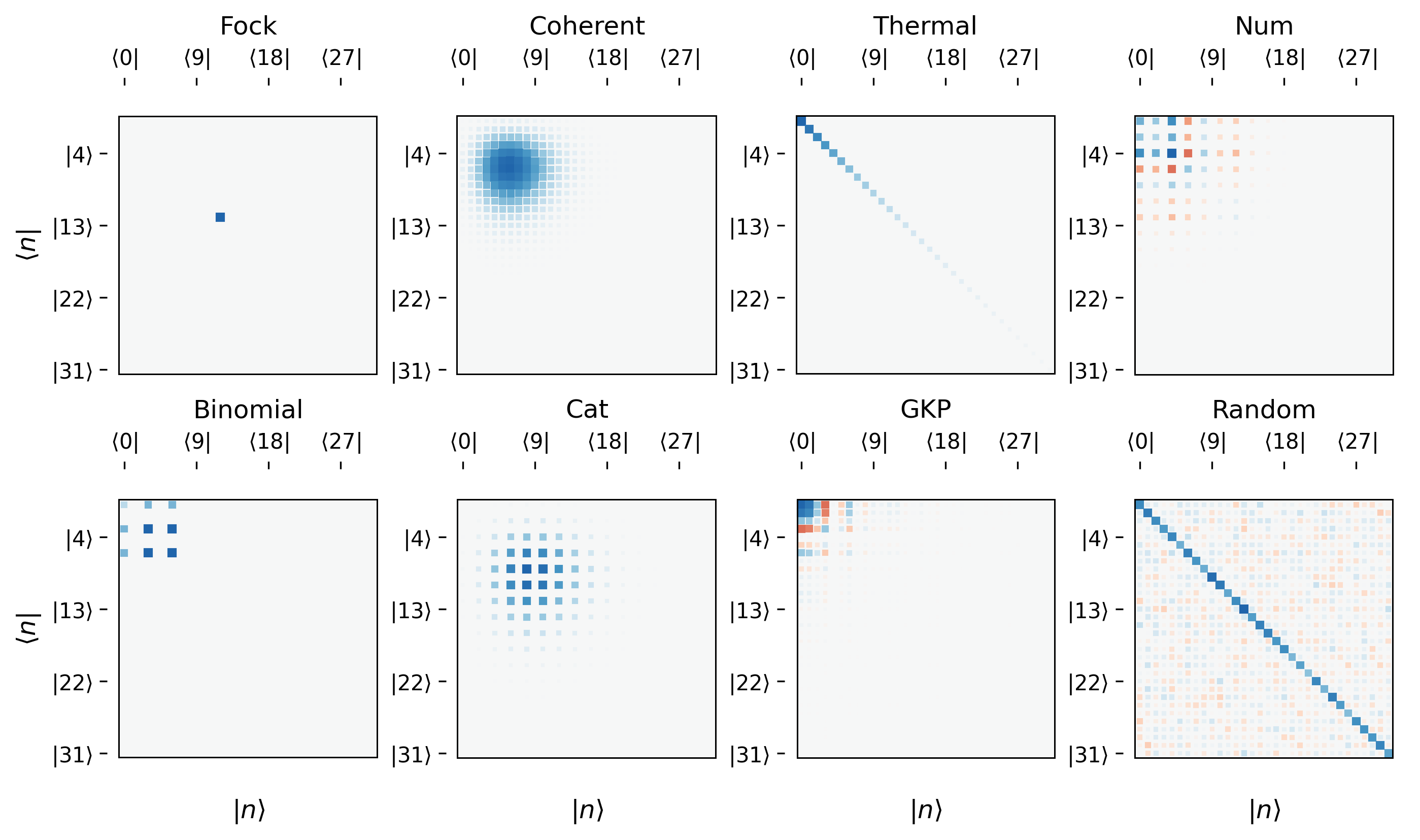}
    \caption{Hinton plots of examples of each of the 8 optical quantum states provided by QSTToolkit.}
    \label{fig:Hintons}
\end{figure}

\subsubsection{Synthetic Measurement Data}
To simulate realistic measurement scenarios, QSTToolkit synthesizes measurement outcomes using two primary regimes: photon occupation number measurements, which mimic number-resolving detector outputs, and Husimi Q function measurements, which emulate displace-and-measure techniques that yield phase space distributions. Given the relevant projective measurement operators, \texttt{qsttoolkit.tomography} could be used for tomography of any optical measurement regime the user desires.

Measurement data is generated according to the Born rule:

\begin{equation}
p_k = \mathrm{Tr} (\rho O_k),
\end{equation} 

where \(O_k\) is the measurement operator associated with measurement outcome \(k\). The right-hand side corresponds to the expected probability for a measurement outcome \(k\), and the sets of expectation values for all outcomes \(k\) are interpreted as the statistical frequency distributions required for reconstruction.

Realistic experimental data are inherently noisy, and QSTToolkit incorporates noise for phase space image data through two primary mechanisms. Firstly, state preparation noise is applied directly to the density matrix, modelled by mixing the pure state \(\rho\) with a random density matrix \(\rho_{rand}\) to simulate thermal interactions with the environment:

\begin{equation}
\rho_{mix} = (1-\zeta)\rho + \zeta \rho_{rand}.
\end{equation}

The second mechanism, measurement noise, is introduced via Gaussian convolution noise (arising from additional bosonic modes from amplification), affine transformations (representing inaccurate phase space displacements for measurement), additive Gaussian noise (modeling the discretization of phase space measurements), and salt-and-pepper noise (pixel over-saturation and non-operation respectively) applied to the measurement images \cite{ahmed2021classification}. These noise models are parametrized, with default values provided in Table~\ref{tab:noise}. These are chosen as they fall within experimentally realistic limits, whilst generally testing tomography models to the limits of their performance. Advanced users can fine-tune noise parameters to match specific experimental conditions.

\begin{table}[htbp]
\caption{Default noise parameters for synthetic data generation.}
\centering
\begin{tabular}{lll}
\toprule
Noise Type & Parameter & Default Value \\
\midrule
Mixed state noise & \(\zeta\) & 0.2 \\
Gaussian convolution & \(n_{th}\) & 2.0 \\
Affine transformation & Rotation \(\theta\) & 20\(^{\circ}\) \\
 & Translation (x, y) & 0.1, 0.1 \\
Additive Gaussian & Std. deviation & 0.01 \\
Salt & Proportion & 0.0 \\
Pepper & Proportion & 0.1 \\
\bottomrule
\end{tabular}
\label{tab:noise}
\end{table}

\begin{figure}
    \centering
    \includegraphics[width=0.6\linewidth]{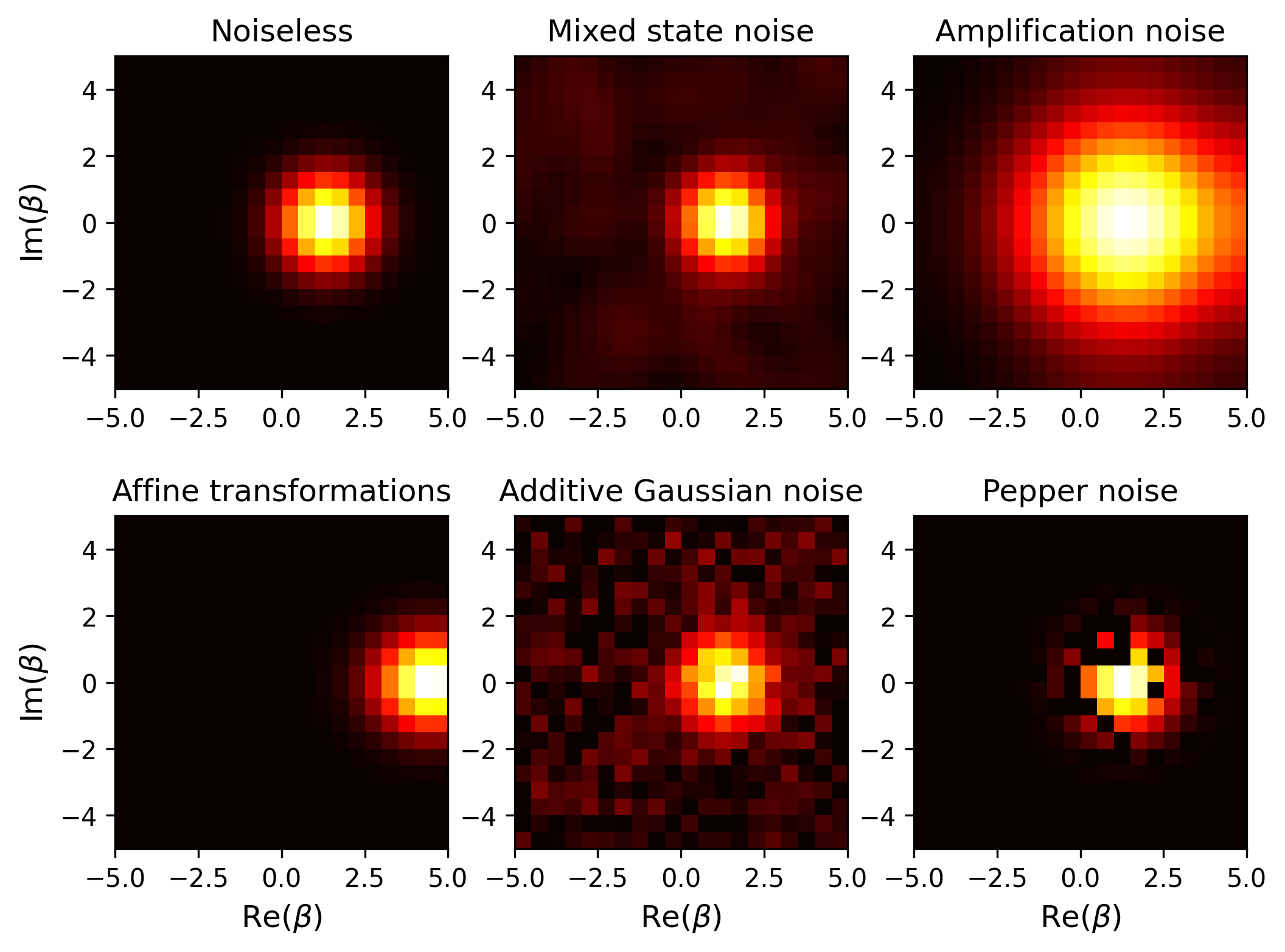}
    \caption{Examples of each noise source applied to a coherent state with \(\alpha = 1.0\). Note that the parameters used in the above are higher than the default values given in Table 2, in order to give exaggerated demonstrations of the effect of each noise source.}
    \label{fig:noise}
\end{figure}

The treatment of noise when approaching tomography, whether the noisy or ‘noiseless’ state is chosen as the target of reconstruction, depends on the type of noise and the use case. In the results in this paper, and in the GitHub example notebooks, the CNN state discrimination and Multitask reconstruction models are trained to identify (and in the case of the latter, reconstruct), noiseless state types from noisy data. The MLE and GAN accurately reconstruct states subject to state preparation (mixed state) noise only. 

For user convenience, a standard training dataset is included via the function \texttt{optical\_state\_dataset(dim, latent\_dim)}. This function generates a collection of 7000 states with both state and measurement noise applied using the parameters in Table~\ref{tab:noise}, enabling standardized comparisons between different model architectures.

\subsection{Tomography Module}
\subsubsection{Overview}
QSTToolkit provides four primary tomography methods: a Maximum Likelihood Estimation (MLE)-based solver which is the standard method \cite{MLE_2, MLE_3, MLE_4} for QST and three deep learning-based models:

\begin{itemize}
    \item A convolutional neural network (CNN) for quantum state discrimination.
    \item A generative adversarial network (GAN) for quantum state reconstruction.
    \item A Multitask hybrid model that simultaneously performs state classification and regression (state characterization).
\end{itemize}

The following sections detail the implemented algorithms, including their theoretical foundations and practical implementations within the toolkit. Where applicable, we provide mathematical formulations to clarify each method’s approach to quantum state reconstruction. Each algorithm is accompanied by example use cases and is supported by recent research demonstrating its effectiveness. The library's GitHub repository\footnote{\url{https://github.com/georgefitzgerald02/qsttoolkit}} features example Jupyter notebooks demonstrating the different methods. 

\subsubsection{Maximum Likelihood Estimation (MLE)}
MLE is a common statistical approach to quantum state tomography that QSTToolkit provides as a baseline method. Using a collection of sampled quadrature values from an unknown optical quantum state, MLE finds the density matrix $\rho$ that maximizes the likelihood of obtaining the observed data. Formally, if $\{O_k\}$ are the measurement operators associated with outcomes $k$ (these satisfy $O_k \ge 0$ and $\sum_k O_k = I$) and $n_k$ is the observed frequency of outcome $k$, the likelihood function (for independent samples) is 

\begin{equation}
    \mathcal{L}(\rho) = \prod_{k} [\Tr(\rho\,O_k)]^{\,n_k},
\end{equation}
or equivalently, the log-likelihood is 
\begin{equation}
    \ell(\rho) = \sum_{k} n_k \ln \Tr(\rho\,O_k).
\end{equation}

The MLE procedure seeks $\rho$ that maximizes $\ell(\rho)$ subject to physicality constraints where $\rho$ must be positive semidefinite ($\rho \succeq 0$) and represent a valid quantum state with a total probability of 1 ($\Tr(\rho)=1$). One way to ensure these two criteria are met is through Cholesky decomposition, where an unconstrained lower triangular matrix $T$ is used to re-parametrize $\rho$ as 

\begin{equation}
    \rho = \frac{T T^\dagger}{\Tr(T T^\dagger)}.
\end{equation}

Reconstructing $\rho$ via this parametrization guarantees that $\rho$ remains positive semidefinite ($\rho \succeq 0$) since $T T^\dagger$ is always Hermitian and positive semidefinite, and the trace normalization ensures that $\Tr(\rho) = 1$, preserving the physicality of the quantum state.


In QSTToolkit, the \texttt{MLEQuantumStateTomography} class uses numerical optimization to execute this constrained maximization, starting from an initial guess \(\rho_0\) and iteratively improving the estimate. The user provides measurement data (which can be generated via the toolkit’s data module or obtained from experiments), and the class will optimize the density matrix. In practice, the algorithm iteratively updates the candidate Cholesky parametrization \(T\) by performing stochastic gradient descent on the resulting \(\ell(\rho)\) to enforce the positive-semidefiniteness and unit-trace conditions at each step. This approach resembles an accelerated projected gradient descent method for the convex optimization problem of likelihood maximization. The implementation relies on TensorFlow’s efficient computation, representing \(\rho\) as a TensorFlow tensor and efficiently computing \(\mathrm{Tr}(\rho\,O_k)\). MLE converges on \(\rho_{\rm MLE}\)—the density matrix that best explains the data in the likelihood sense. The result is stored in the  \texttt{MLEQuantumStateTomography.reconstructed\_dm} attribute, which is a density matrix (outputted as a NumPy array) obtained at the end of optimization. The fidelity, $F$, defined as 

\begin{equation}    
F(\rho,\sigma) = \left(\mathrm{Tr}\sqrt{\sqrt{\rho}\sigma\sqrt{\rho}}\,\right)^2,
\end{equation}

is used to quantify how close this reconstructed state ($\rho$) is to the true underlying state ($\sigma$). It ranges from 0 (no overlap) to 1 (identical states).

MLE is guaranteed to find a physical state and usually yields an excellent estimate given sufficiently many measurements. However, it can be slow to converge for high-dimensional systems or large datasets, since each iteration requires evaluating traces for all outcomes and ensuring positivity. In the optical scenarios targeted by QSTToolkit, the dimension \(N\) (Fock space truncation) can be large (to represent up to, say, 50 photons), making MLE a nontrivial computational task - highlighting the need for the more efficient approaches described later. Regardless, MLE remains the benchmark approach for accuracy given ample data and serves as a reliable reference point for assessing the performance of new methods.

%
%

\subsubsection{CNN-Based Quantum State Discrimination}  
QSTToolkit includes a convolutional neural network (CNN) approach for quantum state discrimination, inspired by recent research that applied deep learning to classify quantum states from experimental-like data~\cite{ahmed2021classification}. The goal of state discrimination differs slightly from full tomography: rather than reconstructing the entire density matrix, it aims to classify the state into one of several known categories or to identify key properties. For example, given a noisy Husimi $Q$ function image of an optical quantum state, a CNN might determine whether the underlying state is a coherent state, a cat state, a thermal state, etc., or identify which basis state is predominant. QSTToolkit’s \texttt{CNNQuantumStateDiscrimination} class is built using TensorFlow's deep learning framework and employs convolutional layers~\cite{sutskeverCNN} that act on 2D arrays (for image-like data such as Husimi $Q$ plots) or 1D sequences (e.g. for photon count distributions), extracting features that are diagnostic of the state. These features are passed through dense layers to output either class probabilities (for discrete classification) or continuous estimates (for regression tasks). The CNN is trained in a supervised manner using a labeled dataset of simulated measurements (typically generated by the toolkit’s data module), learning to map measurement data to the correct state label. 

\begin{figure}
\centering
\includegraphics[width=0.95\linewidth]{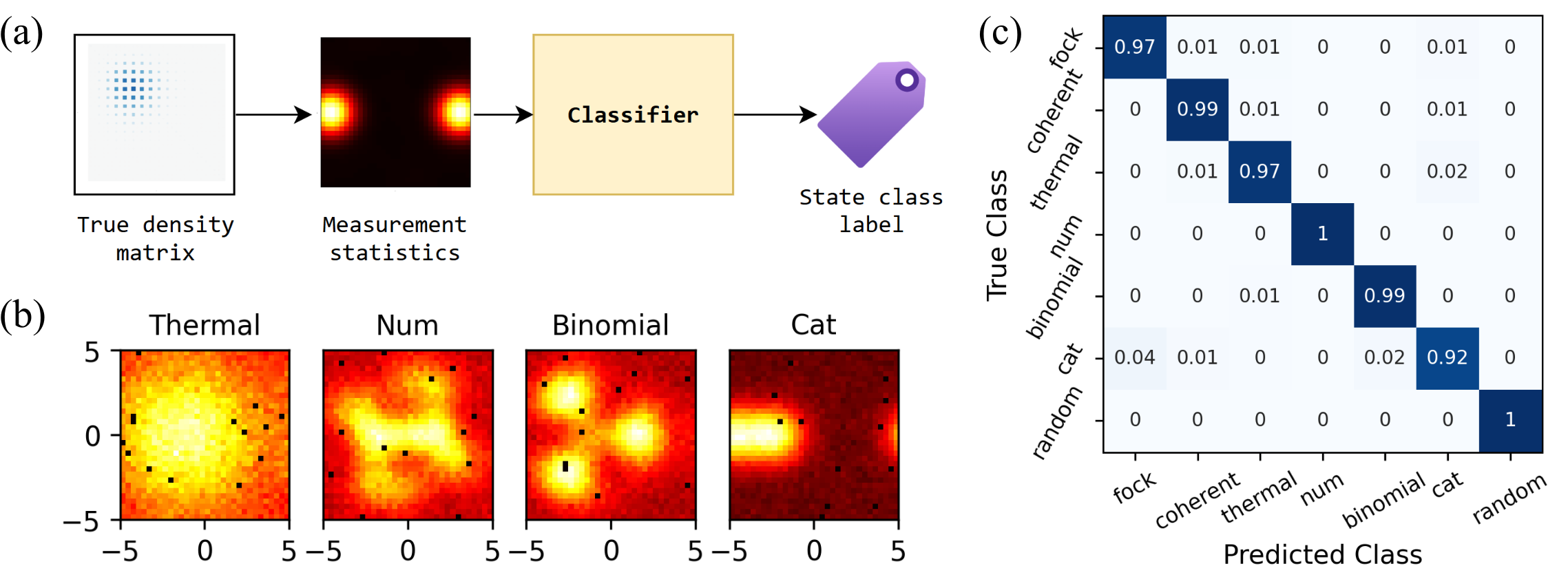}
\caption{A CNN for quantum state discrimination. (a) Schematic of the CNN classifier network operation. (b) Example input Husimi Q function data with state class labels. (c) A classification confusion matrix, demonstrating highly accurate ($>98\%$) state class classification by the CNN network.}
\label{fig:CNN}
\end{figure}

Mathematically, the CNN optimizes a categorical cross-entropy loss
\begin{equation}    
L_{\rm cls} = -\sum_{m} y_m \log \hat{y}_m,
\end{equation}
where $y_m$ is the true label (e.g., state type) and $\hat{y}_m$ is the CNN’s predicted probability for that label, averaged over training examples. Through backpropagation and stochastic gradient descent (handled by TensorFlow), the CNN adjusts its filters to maximize classification accuracy. The general operation of the classifier is conveyed in Figure \ref{fig:CNN}a and the full architecture given in Table~\ref{tab:CNN-arch}. Figure \ref{fig:CNN}c presents classification results by the model when trained to convergence over 31 epochs on the standard dataset of 7000 states provided by QSTToolkit, demonstrating highly accurate (> 98\%) classification. 5600 states are used for the supervised training, and 1400 are set aside for testing, with example input data shown in Figure \ref{fig:CNN}b.

Based on the work of Ahmed et al.~\cite{ahmed2021classification}, the quantum state discriminator is designed to handle noisy, continuous-variable data. Notably, a CNN can identify the most informative regions of the Husimi $Q$ function data (an effect analogous to attention~\cite{attention}), which suggests strategies for adaptive data collection by focusing measurements on regions where the CNN “looks.” In QSTToolkit’s implementation, after training the CNN model (via \texttt{.train()}), it can rapidly classify new measurement data. Since the majority of the compute time is devoted to training, rapid inference is available during experimental conditions requiring real-time state identification. The trained CNN can output results nearly instantaneously on a GPU, whereas methods like MLE might require lengthy iterative computation for each new dataset. Thus, the CNN approach provides a fast, noise-robust discriminator for quantum states, making it ideal for tasks such as monitoring a quantum communication channel or quickly detecting which state a source is emitting from a known set.

\subsubsection{GAN-Based Quantum State Tomography.}
The library includes a Generative Adversarial Network (GAN) for quantum state reconstruction, first introduced in \cite{ahmedCGAN} as a QST-CGAN (quantum state tomography via conditional GAN). It uses two neural networks pitted against each other - a generator $G$ and a discriminator $D$ - to achieve tomography. In this framework, the generator proposes a quantum state given an input vector of measurement statistics, while the discriminator attempts to distinguish between the real measurement data and fake data produced from the generator's proposed state. In practice, the generator outputs a lower triangular Cholesky decomposition (parametrized by network weights $\theta$) and reconstructs a density matrix $\rho_G$ as in equation (6) in order to ensure the same physicality conditions as in MLE. QSTToolkit then simulates the measurement process on $\rho_G$ by calculating expectation values for given measurement operators, before feeding this simulated $\text{data}_{G}$ into the discriminator.

\begin{figure}[ht]
    \centering
    \includegraphics[width=0.75\linewidth]{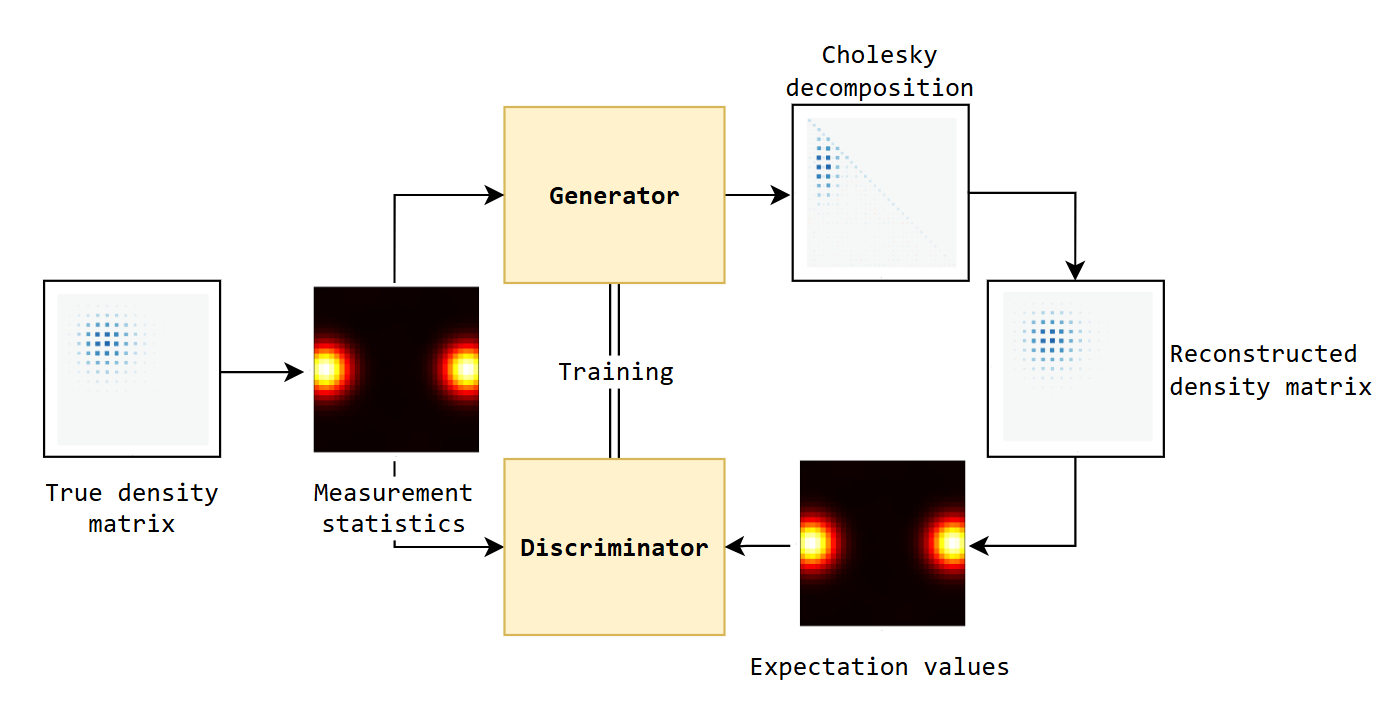}
    \caption{Schematic of the GAN QST network operation. The generator network takes measurement statistics as an input, and produces a Cholesky decomposition of its proposed state. This is converted into a density matrix, from which measurement expectation values are calculated. The discriminator trains to distinguish between the original and reconstructed states' data.}
    \label{fig:GAN_diagram}
\end{figure}

The discriminator is a classifier that learns to output ``real'' if the data originated from the actual experiment and ``fake'' if it originated from $G$. The generator is trained to fool the discriminator - that is, to produce states whose simulated data cannot be distinguished from the real data $\text{data}_{G}$ - while the discriminator is simultaneously trained to improve its discrimination. This adversarial training continues until an equilibrium is reached, where the generator's state produces data nearly identical to the real measurements, at which point the generator's state should be an excellent reconstruction of the true state.

At each training step, the GAN training optimizes the discriminator's binary cross-entropy loss

\begin{equation}
L_D = -\Bigl( \ln D(\text{data}_{\rm real}) + \ln\bigl[1 - D(\text{data}_{G})\bigr]\Bigr),
\end{equation}

and the generator's loss, which can be taken as

\begin{equation}
L_G = -\ln D(\text{data}_{G}),
\end{equation}

thereby encouraging $G$ to increase the discriminator's error rate. In a QST-CGAN, the standard GAN is modified by making the discriminator itself a ``learnable loss'' function for the generator. Instead of pre-defining a metric (such as the mean square error) between the real and simulated data, the discriminator learns the most pertinent features to differentiate the data. According to the results in \cite{ahmedCGAN} (replicated by QSTToolkit in Section~\ref{results}), the GAN-based tomography can reconstruct quantum states using up to two orders of magnitude fewer iterative steps than iterative MLE methods, and requires far fewer samples of data to reach a given fidelity. This dramatic improvement in epoch efficiency is a key advantage of the GAN approach.

Within QSTToolkit, the class \texttt{GANQuantumStateTomography} implements this adversarial training. The user provides the measured data of an unknown state, and the model then initializes a generator network (which is, by default, a deep neural network that outputs a Cholesky decomposition of $\rho$) and a discriminator network. It runs the training loop (which can be computationally intensive, but harnesses GPUs via TensorFlow) until convergence. The output is a reconstructed density matrix, $\rho_{\rm GAN}$, which can be accessed similarly to the MLE case (as \texttt{GANQuantumStateTomography.reconstructed\_state}). This module uses QuTiP to help turn a candidate state into expected measurement outcomes efficiently, and employs TensorFlow for the neural networks and training process.
Currently \texttt{GANQuantumStateTomography} supports reconstruction of individual density matrices from a single input vector. Future work could investigate transfer learning (using a GAN trained to reconstruct one state, to reconstruct a different state) using the architecture, or tomography of batches of states at once in the same training.

\subsubsection{Multitask Neural-Network Tomography.}  
Building upon the work of Luu et al.~\cite{luu2024universal}, QSTToolkit includes a Multitask learning approach to quantum state characterization based on a network that simultaneously performs several estimation tasks. This approach is motivated by the concept of “universal” quantum tomography with deep neural networks - designing a single model that can handle a wide variety of states (pure or mixed) and output both discrete and continuous information about the state. The model takes input measurement data and produces both a classification output (e.g., identifying the state type or category) and a regression output (predicting a key parameter in the initialization of the state using QuTiP). For instance, a user could train the Multitask network to classify a given state as a cat state, a Fock state, or a coherent state, while simultaneously estimating the exact photon number (if it is a Fock state) or the displacement amplitude (if it is a coherent or cat state). The network typically employs a shared feature-extraction "trunk" (e.g., convolutional layers processing the input data), which then splits into two “branches” - one for producing class probabilities and another for generating continuous values. The loss function for training is a weighted sum of the classification loss and regression loss, ensuring that the network balances both tasks effectively. By training on a diverse dataset (covering many state types and noise levels), the Multitask model aims to generalize across the optical quantum state space, thereby realizing the vision of “universal” tomography.

\begin{figure}
    \centering
    \includegraphics[width=1\linewidth]{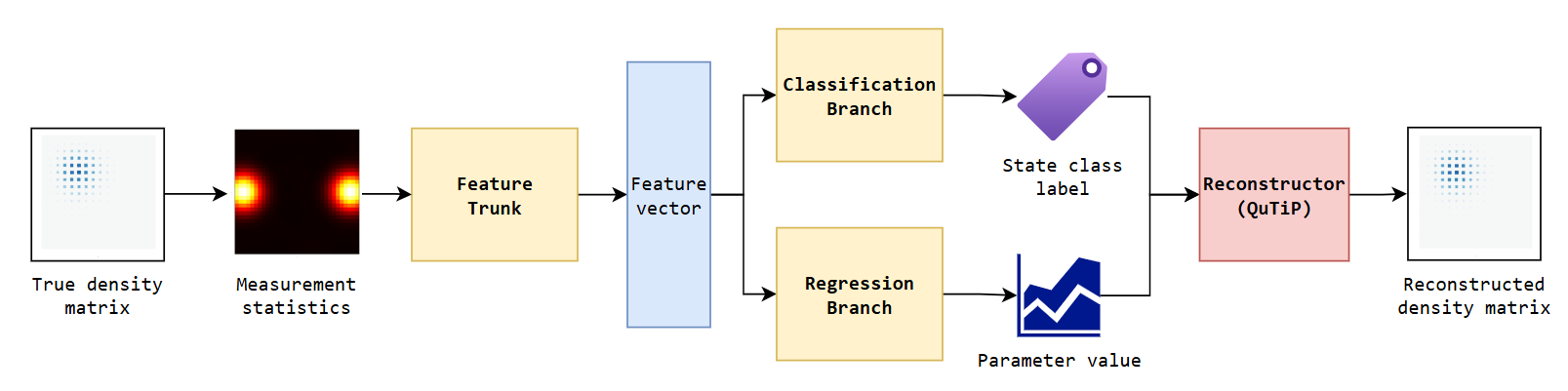}
    \caption{The Multitask state reconstructor network. Measurement statistics enter the feature trunk, which outputs a "feature vector" containing the necessary information about the state. This passes into a classification and a regression branch, which infer a state class label and a key parameter value respectively. These are used by the reconstructor to synthesize an output density matrix.}
    \label{fig:Multitask_diagram}
\end{figure}

A key benefit of the Multitask approach is improved efficiency: the classification task can help the network learn high-level distinctions between states, which in turn provides context for the regression task to fine-tune parameters, and vice versa. This cross-task information sharing often leads to better performance on each individual task than training two separate networks (one for classification and one for parameter estimation). By knowing the state class, the network can apply the appropriate interpretation to the regression output. For example, if the network classifies the state as a “GKP state”, it can infer that the regression output may correspond to an overall squeezing parameter or displacement error. 

\begin{figure}[h!]
    \centering
    \includegraphics[width=0.485\linewidth]{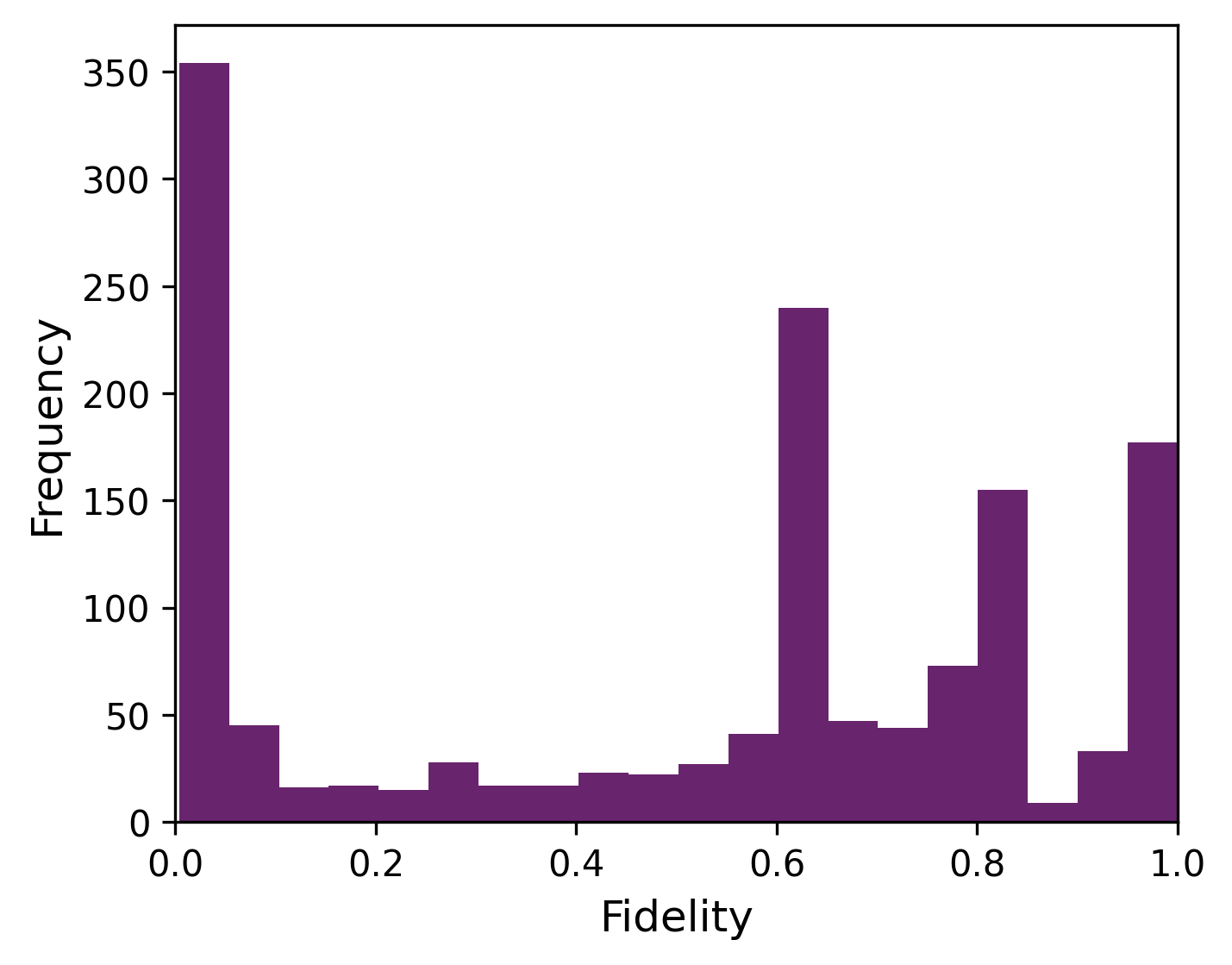}
    \caption{Distribution of fidelities between the test set's true and reconstructed states by the Multitask network. Results are disjointed, with reconstructions generally either very accurate or significantly far off.}
    \label{fig:fidelities}
\end{figure}

In QSTToolkit, a Multitask neural network is implemented using TensorFlow in the \texttt{MultitaskQuantumStateTomography} class, and a specialized \texttt{StateReconstructor} class is provided to synthesize optical quantum states based on the model's inferences. Unlike other models where physical reconstructions are ensured by a Cholesky decomposition, the regression head is restricted to output only physically allowed values for a given predicted class, and a combination of QuTiP and QSTToolkit functions are employed to generate the reconstructions. Figure \ref{fig:fidelities} illustrates the results of reconstruction of the model trained to convergence over 125 epochs on the QSTToolkit standard dataset train/test split, giving an average reconstruction fidelity of 0.500. The results are disjointed, with predictions generally either very accurate or inaccurate - for many state types, even a slight discrepancy in the inferred regression parameter can lead to a drastically different reconstructed density matrix and a low reconstruction fidelity. Nevertheless, the high accuracy of the individual state classification and regression by the respective branches makes this a useful approach for detailed and efficient state characterization, even if the performance of the full reconstruction is limited.

\newpage
\section{Illustrative Examples}
\label{results}
\subsection{Example usage}
The following example demonstrates how QSTToolkit can be used to perform quantum state tomography with a Generative Adversarial Network (GAN). In this scenario, the toolkit trains its default GAN architecture to reconstruct the density matrix of a num state from Husimi Q function phase space measurement data. This illustrates QSTToolkit's streamlined capabilities, from dataset creation to neural network training and performance evaluation, providing a user-friendly entry point into advanced quantum tomography methods.

\begin{tcolorbox}[colback=gray!5!white,colframe=gray!75!black,title=GAN Tomography Example]
\begin{verbatim}
import numpy as np
import qsttoolkit as qst

# Prepare state (e.g. a num state)
dim = 32             # Hilbert space dimensionality
num_test = qst.num_dm(`M2', dim)

# Prepare measurement data
data_dim = 20        # Phase space grid dimensions
xgrid = np.linspace(-5, 5, data_dim)
pgrid = np.linspace(-5, 5, data_dim)
measurement_operators = qst.tomography.measurement_operators(dim, `Husimi-Q',
                                                            xgrid=xgrid, pgrid=pgrid)
expectation_values = qst.expectation(num_test, measurement_operators)
measurement_data = expectation_values.numpy().reshape(1, data_dim**2)

# GAN tomography
GAN_reconstructor = qst.tomography.GANQuantumStateTomography(data_dim=data_dim**2)
GAN_reconstructor.reconstruct(measurement_data, measurement_operators, epochs=1000)
result = GAN_reconstructor.reconstructed_dm
\end{verbatim}
\end{tcolorbox}

\subsection{Model Comparison}
\begin{figure}[!ht]
    \centering
    \includegraphics[width=0.5\linewidth]{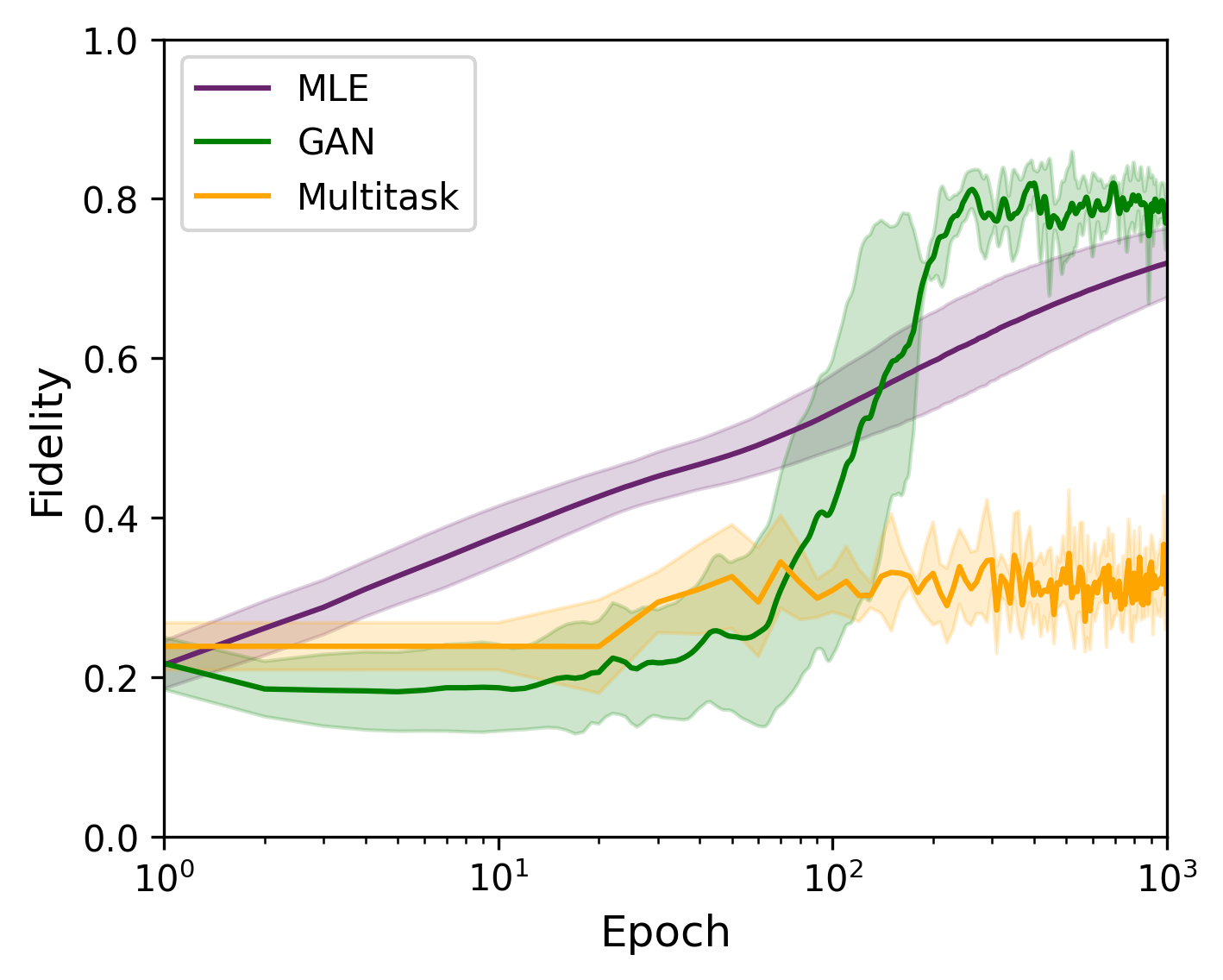}
    \caption{Comparison of the performance of the three QST models - average reconstructed state fidelities over 1000 training epochs with standard deviation error regions (left) and the corresponding training time (right) across 5 reconstruction runs. The MLE and GAN train to reconstruct the density matrix of the same num state (type `M2') subject to mixed state noise ($\zeta=0.2$). The Multitask model trains on a dataset of 5600 states of a variety of types with the same noise level applied, and an average reconstruction fidelity for 200 test num states are calculated every 10 training epochs.}
    \label{fig:comparison}
\end{figure}

An example of the tomography performance comparison made possible by QSTToolkit is illustrated in Figure \ref{fig:comparison}. Three models - Maximum Likelihood Estimation (MLE), a Generative Adversarial Network (GAN), and Multitask learning - are compared based on their reconstruction fidelity over 1000 training epochs, averaged across 5 reconstructions of num states. The MLE and GAN are trained to reconstruct the density matrix of the same single num state (type `M2') subject to mixed state noise ($\zeta=0.2$). After 1000 training epochs, the MLE models reach an average fidelity of 0.719. The GAN model demonstrates notable fidelity improvement after approximately 100 epochs, achieving the highest 1000-epoch average fidelity (0.771) among the models tested. The Multitask model trains on a dataset of 5600 states of a variety of types with the same noise level applied, and an average reconstruction fidelity for 200 test num states are calculated every 10 training epochs. This is repeated for 5 instances of the model, and the fidelities are averaged to allow for a valid performance comparison with the single state reconstruction models. The final Multitask model reconstructs the test states to an average fidelity of 0.305. The GAN and Multitask models are trained using an NVIDIA A100 GPU with 40GB RAM, and the MLE is trained using an Intel i5 4-core CPU with 8GB RAM.

This comparison highlights the value QSTToolkit provides for comparing model performance in different situations - in the case of num states with low-resolution, noisy data, their complex Husimi Q image patterns are "learnt" by the GAN in fewer training epochs than the MLE, albeit over a longer time period. The Multitask model struggles to accurately reconstruct num states since the inferred parameter by the regression branch (mean photon number) is discretized to physically allowed values by the state reconstructor. A prospective user of the toolkit may wish to run a similar simulation based on their particular experimental scenario and available computational resources to ascertain which model is the best option for their use case.

\section{Conclusion} QSTToolkit provides a comprehensive Python framework that combines synthetic data generation, noise modelling, and quantum state tomography using both traditional statistical and modern deep learning methods. Designed with optical quantum states in mind, it is particularly suited for research in quantum optics, quantum error correction, and quantum communication, where efficient and accurate state reconstruction is critical. Its deep learning models enable real-time monitoring and classification, making it well-suited for dynamic experimental conditions where stability is a concern. For example, QSTToolkit’s CNN-based classifiers can be integrated into continuous variable quantum key distribution (CV-QKD) setups to quickly detect deviations in state preparation - since the computationally intensive training is performed only once, rapid inference is available during operation.

Furthermore, QSTToolkit offers an end-to-end pipeline - from simulation to reconstruction to validation - that simplifies the workflow for quantum state characterization. This integration could help accelerate research and provide a tool for theoretical investigations and algorithm development, as its standardized datasets and benchmarking capabilities allow researchers to compare new tomography approaches on equal footing. Additionally, the toolkit is well-suited for educational use, offering accessible visualization tools and pre-built examples that lower the barrier for students to experiment with both traditional and deep-learning-based methods.



\bibliographystyle{unsrt}  
\bibliography{references}

\newpage

\appendix

\section{Availability of QSTToolkit Online}
QSTToolkit is available for download via the PyPI package manager. Detailed installation instructions and the full source code are hosted at:
\begin{center}
  \url{https://pypi.org/project/qsttoolkit/} \\
  \url{https://github.com/georgefitzgerald02/qsttoolkit}.
\end{center}

Comprehensive documentation of the full range of functions and classes provided by QSTToolkit can be found at:
\begin{center}
  \url{https://qsttoolkit.readthedocs.io/en/latest/}.
\end{center}

\section{Model Architectures}
For completeness, we include key architectural details of the implemented models.

\subsection{CNN-based quantum state discrimination model}
The CNN classification model employs sequential convolutional layers and max pooling to extract spatial features, followed by dense layers to classify quantum states into one of seven categories, as detailed in Table~\ref{tab:CNN-arch}.

\begin{table}[htbp]
\caption{Model architecture for the \texttt{CNNQuantumStateDiscrimination} model.}
\centering
\begin{tabular}{ll}
\toprule
Layer & Activation \\
\midrule
Conv2D (32 filters, 3x3) & ReLU \\
MaxPooling2D & -- \\
Conv2D (64 filters, 3x3) & ReLU \\
MaxPooling2D & -- \\
Conv2D (128 filters, 3x3) & ReLU \\
MaxPooling2D & -- \\
Flatten & -- \\
Dense (128 units) & ReLU \\
Dense (7 units) & Softmax \\
\bottomrule
\end{tabular}
\label{tab:CNN-arch}
\end{table}

\subsection{Multitask Quantum State Tomography}
The Multitask model begins with a shared convolutional `feature trunk' (Table~\ref{tab:MT-branch}) followed by two separate heads for classification (Table~\ref{tab:MT-class}) and regression (Table~\ref{tab:MT-reg}).

\begin{table}[H]
\caption{Initial branch architecture for the MultitaskQuantumStateTomography model.}
\centering
\begin{tabular}{ll}
\toprule
Layer & Activation \\
\midrule
Conv2D (32 filters, 3x3) & LeakyReLU \\
Conv2D (32 filters, 3x3) & LeakyReLU \\
GaussianNoise (0.1) & -- \\
Dropout (0.2) & -- \\
Conv2D (64 filters, 3x3) & LeakyReLU \\
Conv2D (128 filters, 3x3) & LeakyReLU \\
GaussianNoise (0.1) & -- \\
Dropout (0.2) & -- \\
Conv2D (256 filters, 3x3) & LeakyReLU \\
Conv2D (512 filters, 3x3) & LeakyReLU \\
Dropout (0.2) & -- \\
Flatten & -- \\
\bottomrule
\end{tabular}
\label{tab:MT-branch}
\end{table}

\begin{table}[H]
\caption{Classification branch for the MultitaskQuantumStateTomography model.}
\centering
\begin{tabular}{ll}
\toprule
Layer & Activation \\
\midrule
Dense (64 units) & LeakyReLU \\
Dropout (0.2) & -- \\
Dense (128 units) & LeakyReLU \\
Dense (7 units) & Softmax \\
\bottomrule
\end{tabular}
\label{tab:MT-class}
\end{table}

\begin{table}[H]
\caption{Regression branch for the MultitaskQuantumStateTomography model.}
\centering
\begin{tabular}{ll}
\toprule
Layer & Activation \\
\midrule
Dense (64 units) & LeakyReLU \\
Dense (128 units) & LeakyReLU \\
Dense (256 units) & LeakyReLU \\
Dense (256 units) & LeakyReLU \\
Dense (256 units) & LeakyReLU \\
Dense (128 units) & LeakyReLU \\
Dense (2 units) & -- \\
\bottomrule
\end{tabular}
\label{tab:MT-reg}
\end{table}

\subsection{GANQuantumStateTomography Model}
The GAN framework consists of a generator (Table~\ref{tab:GAN-gen}) and a discriminator (Table~\ref{tab:GAN-disc}) network.

\begin{table}[H]
\caption{Generator network for GANQuantumStateTomography.}
\centering
\begin{tabular}{ll}
\toprule
Layer & Activation \\
\midrule
Dense (512 units) & LeakyReLU \\
Reshape (16x16x2) & -- \\
Conv2DTranspose (64 filters, 4x4, stride 2) & LeakyReLU \\
BatchNormalization & -- \\
Conv2DTranspose (64 filters, 4x4, stride 1) & LeakyReLU \\
BatchNormalization & -- \\
Conv2DTranspose (32 filters, 4x4, stride 1) & LeakyReLU \\
Conv2DTranspose (2 filters, 4x4, stride 1) & -- \\
CholeskyLowerTriangular (custom) & -- \\
\bottomrule
\end{tabular}
\label{tab:GAN-gen}
\end{table}

\begin{table}[H]
\caption{Discriminator network for GANQuantumStateTomography.}
\centering
\begin{tabular}{ll}
\toprule
Layer & Activation \\
\midrule
Dense (128 units) & LeakyReLU \\
Dense (64 units) & LeakyReLU \\
Dense (32 units) & LeakyReLU \\
Dense (1 unit) & Sigmoid \\
\bottomrule
\end{tabular}
\label{tab:GAN-disc}
\end{table}

%
%
%
%
%
%
%
\end{document}